\documentclass[oldversion]{aa}
\usepackage{epsfig}
\usepackage{graphicx}

\begin{document}

\title{Velocity dispersion around ellipticals in MOND}

   \author{O. Tiret         \inst{1}           
\and          F. Combes         \inst{1}  
\and G. W. Angus   \inst{2}           
\and B. Famaey   \inst{3} 
\and H.S. Zhao \inst{2} 
 }
  \offprints{O. Tiret }

   \institute{ Observatoire de Paris, LERMA,    61 Av. de l'Observatoire, F-75014, Paris, France  
\and  SUPA, School of Physics and Astronomy, Univ. of St. Andrews, Scotland KY169SS
\and Institut d'Astronomie et d'Astrophysique, Univ. Libre de Bruxelles, CP 226, Bvd du Triomphe, B-1050, Bruxelles, Belgium
 }

   \date{Received XXX 2007/ Accepted YYY 2007}

   \titlerunning{Velocity dispersion around ellipticals in MOND}

   \authorrunning{Tiret et al.}

\abstract{We investigate how different models that have been proposed for solving the dark matter 
problem can fit the velocity dispersion observed around elliptical galaxies, on either a small scale
($\sim$ 20kpc) with stellar tracers, such as planetary nebulae, or large scale
($\sim$ 200kpc) with satellite galaxies as tracers. 
Predictions of Newtonian gravity, either containing pure baryonic
matter, or embedded in massive cold dark matter (CDM) haloes, are compared with predictions
of the modified gravity of MOND.
   The standard CDM model has problems on a small scale, and the Newtonian pure baryonic model
has difficulties on a large scale, while a fit with MOND is possible on both scales.
}
\maketitle


\section{Introduction} 

Measuring the velocity field in and  around galaxies is the main
way to test the dark matter distribution at small and intermediate scales.
  The observation of what are apparently non-Newtonian rotation curves around spiral galaxies
(e.g. Rubin et al. 1980) has been first solved by assuming that
galaxies are embedded in dark matter haloes. Numerical simulations, however, predict a radial distribution for the CDM model much more concentrated than what is observed (Gentile et al 2004; de Blok 2005).

An alternative explanation was proposed by Milgrom (1983),
as MOdified Newtonian Dynamics (MOND).
When the Newtonian acceleration 
falls below the critical value  $a_0 \sim 2 \times 10^{-10}$ m s$^{-2}$,
the gravity law is empirically modified and then declines in 1/$r$
instead of 1/r$^2$. Around spherical systems, the modified acceleration $g$ 
satisfies the relation
$$ g \mu(g/a_0)  =  g_n$$
where $g_n$ is the Newtonian acceleration. For non-spherical geometry, 
this is only an approximation. However, in this Letter we only consider spherical systems as representing elliptical galaxies and we adopt $\mu(x) = x/\sqrt{1+x^2}$.

This model is very successful on a galactic scale; in particular, it explains a large number of rotation curves of galaxies, with some exceptions (Gentile et al. 2004), and naturally the Tully-Fisher relation,
(e.g. Sanders \& McGaugh 2002). While the dark matter problem is observationally very clear around
spiral galaxies, thanks to their rotation curve measured  with the cold 
hydrogen gas at 21cm, which is nearly in circular orbits (e.g. Bosma 1981, 
Verheijen \& Sancisi 2001), the situation is much more complex around
elliptical galaxies, with little or no rotation. 

Recently, planetary nebulae have been used as an efficient tool for measuring the velocity field at 
large radii in early-type galaxies, and they complement stellar absorption kinematical studies 
(Romanowky et al. 2003). In typical elliptical galaxies, the velocity-dispersion profiles were found to decline with radius, up to 5 effective radii, thereby requiring no dark matter at all. Dekel et al.
 (2005) show that the data are still compatible with the usual dark matter models, if the 
planetary nebulae tracers have particularly radial orbits in the outer parts, because of a recent 
merger with small impact parameter. However, the recent results from Douglas et al (2007) 
challenge this interpretation, prolonging the decline to more than 7 effective radii.

On larger scales around early-type galaxies, from 50 to 300kpc,
Klypin \& Prada (2007), KP07, have proposed to test gravity models with satellite
galaxies as a tracer.  From the Sloan Digital Sky Survey, they stack several
thousand galaxies in 3 luminosity classes and determine the number
density of satellites and their velocity dispersion around them. In each mass
range, the radial distributions are obtained with around 1500 satellites, although
about 1.5 satellite exist around each galaxy.

This large-scale galaxy neighborhood has not been widely tested yet
in modified gravity. The well-known difficulty of MOND in clusters has
found a possible solution with neutrinos of  2 eV mass (Sanders 2003; 
Angus et al 2007a), and the escape velocity around giant galaxies 
like the Milky Way was shown to correspond to observations, when 
including the external field effect (Famaey et al 2007; Wu et al 2007).

In this work, we  solve the Jeans equation for the distribution
of the velocity dispersion around elliptical galaxies, and in particular we fit the NGC3379 galaxy, where the most extended data is 
available for the velocities.
We also further explore fits at larger radii for the special case of
NGC 3379, with satellite galaxies 
as tracers, as done by KP07 and
Angus et al (2007b). This is only statistically valid around a generic early-type galaxy, 
with a mass comparable to NGC 3379. This is the brightest galaxy of a group, but the observed companion 
velocity is not statistically significant.

\begin{figure}[ht!]
\center{
 \includegraphics[angle=-90, width=8cm]{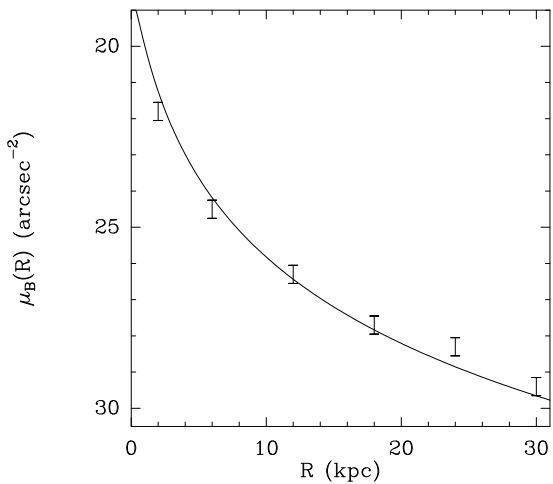}
\caption{ Fit of the blue-light distribution in NGC 3379.
The data points  are taken from Douglas et al (2007). The curve represents the light profile derived with the mass-to-light ratio given in Table \ref{massp} and a mass model of cored-Hernquist type (Table \ref{tracers}).
}
\label{logsigma}
}
\end{figure}

 \vspace{-1 cm}

\section{The Jeans equation}

Because of the spherical symmetry, it has been shown (Angus et al. 2007) that the Jeans equation can be written in MOND, as well as in DM:
$$ \frac{d \sigma^2}{dr} + \sigma^2 \frac{(2 \beta + \alpha)}{r} = - g(r) $$
where $\sigma$ is the radial velocity dispersion, $\alpha =  d ln \rho / d ln r$ is the slope of the tracer density $\rho$,
and $\beta = 1-  (\sigma_\theta ^2 +  \sigma_\phi ^2) / 2 \sigma^2$ 
 is the velocity anisotropy.

The Jeans equation is usually used for a unique self-gravitating component,
where the density $\rho$ appearing in the ``pressure'' term on the left is the 
same as the density appearing in the Poisson equation, giving the density field
$g(r)$ on the right. On the left side, we use the density and the velocity dispersion
of the tracers only, which can be very different from the density producing the
potential, in particular for the satellite galaxies, which act as test particles.

In the present approach, we want to compare all models (Newtonian, with or without
CDM and MOND) with the same dynamics for the Jeans equation, fitting the
density of tracers as close as possible to the observed density distribution.
 For the CDM models, we consider two different amounts of dark matter:
(i) the CDM1 model, reminiscent of what is found in cosmological
simulations, where the amount of dark matter inside the radius of 200kpc
is equal to 60 times the visible matter; (ii) the CDM2 model, more
akin to what is required to model rotation curves of spiral galaxies:
the amount of dark matter inside the radius of 200kpc
is equal to 20 times the visible matter.

  All models are then compared with the same tools, while KP07
compare the CDM model with numerical simulations, and MOND with the Jeans equation, but without exploring all variations of the
tracer density and anisotropy profiles.

\begin{figure*}[ht!]
\center{
\includegraphics[width=17cm,angle=0]{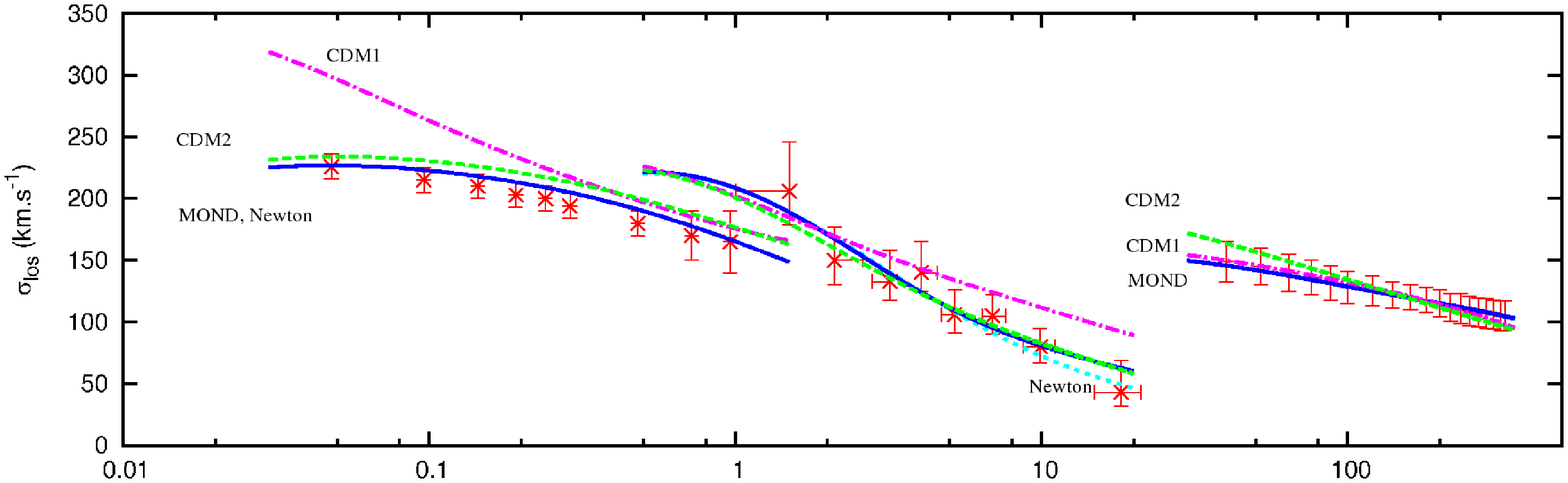}
\includegraphics[width=17cm,angle=0]{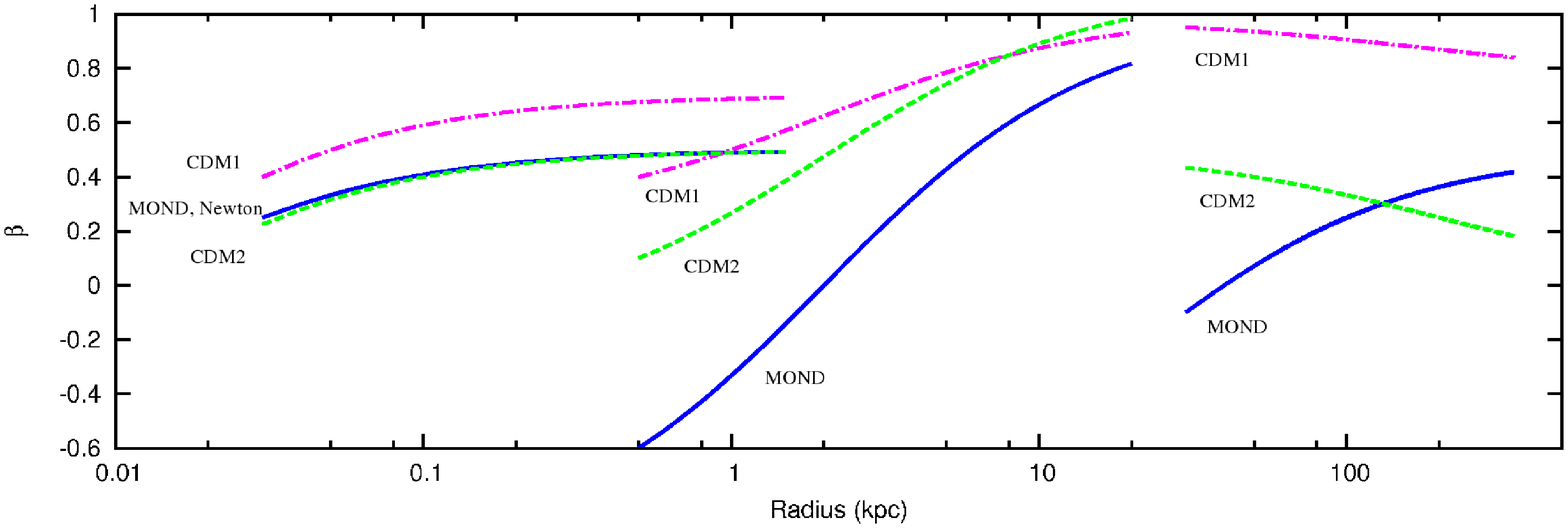}

\caption{ Fit of the velocity dispersion data around NGC 3379
with pure a Newton model without DM, CDM1, CDM2, and MOND.
The data points in the very center ($r<1$ kpc) are from Shapiro
 et al. (2006). On intermediate scales they are taken from Douglas et al. (2007)
and are based around 10-30kpc on the planetary nebulae as tracers.
  On large scales, the data points correspond to the statistical distribution from
the satellite galaxies as tracers (KP07),
adapting NGC 3379 in their small luminosity class.
 The tracer density is taken as an Hernquist distribution
on a small scale, and fitted to the satellite density distribution
on a large-scale.  
 The best fit is obtained with varying $\beta(r)$. Purple/dot-dash: CDM1.
Green/dash: CDM2.  Blue/full line: MOND, Cyan/short dash: Newton without DM.
}
\label{bestfit}
}
\end{figure*}

\begin{table}
\caption{Parameters for deriving the mass of NGC 3379.}
\begin{tabular}{ c c c c}
\hline\hline
  D (Mpc) &  M$_B$  & B$_{Tc}$  & M/L$_B$  \\
\hline
  9.8   &  -19.9   &  10.10   & 7.0    \\
\hline
\end{tabular}
\label{massp}
\end{table}

\section{Results}

\begin{table}
\caption{Radial distribution of the tracer density}
\begin{tabular}{c c c c c}
\hline\hline
& $\alpha_0$& $\alpha_1$& $r_\alpha$ & $\epsilon$  \\
\hline
stars \& PN    & $-1$ &$ -3$    & $1.1$ & $1.2$     \\
\hline
Sat    & $-1.7$ &$-1.5$    & $150$&  $0$  \\
\hline
\end{tabular}
\\
\label{tracers}
$\rho(r)\propto (r+\epsilon)^{\alpha_0}(r+r_\alpha)^{\alpha_1}$ ;  
 $\epsilon$, $r_\alpha$ in kpc \\

\end{table} 

NGC 3379 is modeled by a cored-Hernquist distribution (see Fig. \ref{logsigma}). The stellar 
mass of this galaxy is estimated to be $M=1 \times10^{11} M_\odot$, assuming a constant
mass-to-light ratio M/L  (see Table \ref{massp}).
In MOND (and Newton without DM), this stellar distribution  is the only matter contributing to the 
gravitational potential. In Newtonian dynamics with DM, the DM halo follows the NFW radial profile found 
in N-body cosmological simulations. We used three kinds of tracers to compare Newton without DM, CDM1,
 CDM2, and MOND models, for different length scales:  stars for the inner part ($r < 1$ kpc, Shapiro
 et al 2006), planetary nebulae for the middle part ($1$ kpc $< r<20 $ kpc, Douglas et al. 2007), 
and satellites ($20$ kpc $< r< 200 $ kpc, KP07). The satellites do not correspond to the real ones 
orbiting NGC 3379, but are a statistical representation from the SDSS (KP07), giving the velocity dispersion 
of satellites submitted to the gravity of a typical elliptical galaxy in the same mass category as NGC 3379.
  The parameters of the tracer density are displayed in Table \ref{tracers}.

We numerically solve the Jeans equation for each tracer. The projected velocity dispersion 
($\sigma_{los}$) is represented in Fig. \ref{bestfit}.

\subsection{Newton dynamics without DM}
\begin{itemize}
 \item Stars:
the projected velocity dispersion ($\sigma_{los}^\star$) is well-fitted using a small radial
 anisotropy increasing with radius from $\beta=0.2$ to $\beta=0.4$.
Since the acceleration here is above the critical $a_0$, both MOND and Newton share
the same fitting parameters.
 \item PN:
as for the stars, the observed PN velocity dispersion ($\sigma_{los}^{PN}$) is also in good 
agreement with the Newtonian model. It implies a tangential velocity in the central region 
($\beta=-0.6$) evolving to radial trajectories ($\beta=0.8$). 

\item Satellites:
$\sigma_{los}^{sat}$ cannot be fitted at all by considering the stellar mass alone. 
A dark matter component is required.
\end{itemize}
\vspace{-0.5 cm}
\subsection{CDM1: DM halo mass from cosmological simulations}

The mass of the NFW halo is 60 times the stellar mass inside $200 kpc$.

\begin{itemize}
 \item Stars:
the cusp of the DM halo does not allow fitting the $\sigma_{los}^\star$, whatever the anisotropy
 profile. The predicted velocity dispersion is too large.
 \item PN:
$\sigma_{los}^{PN}$ does not fit the entire data set. The dispersion is too large around 
$10$ kpc. If $\beta$ is increased towards its maximum, 
up to a purely radial anisotropy profile (which tends to decrease 
the velocity dispersion), $\sigma_{los}^{PN}$ is then too small in the inner part. In the best fit, 
 the value of $\beta$ is maximum at $1$ kpc.
 \item Satellites:
the model reproduces the observations well, if radial anisotropy is maintained
all over the region ($\beta > 0.8$).
\end{itemize}
\vspace{-0.5 cm}
\subsection{CDM2:  DM halo mass compatible with average spirals}

The mass of the NFW halo is 20 times the stellar mass inside $200 kpc$.
\begin{itemize}
 \item Stars \& PN:
they are well-fitted, and the DM halo cusp is no longer a problem in the internal dynamics of the 
 galaxy. For the PN, the velocity dispersion is isotropic near the center ($r<1$ kpc) 
and increase to its maximum $\beta = 1$ at $20$ kpc.
 \item Satellites:
it is more difficult to fit $\sigma_{los}^{sat}$ with less DM. We need to impose an isotropic 
velocity dispersion  ($\beta \sim 0$), to avoid too fast 0a fall in $\sigma_{los}^{sat}$.
The negative gradient of $\beta (r)$ helps to straighten the slope of 
$\sigma_{los}^{sat}(r)$ which otherwise is too steep.
\end{itemize}

We also varied the dark-to-visible mass ratio between $10$ to $60$, and $20$ gives the best compromise between the $1-10$ kpc and $30-200$ kpc regions.

\begin{table}
\caption{ Best fit for the  anisotropy distributions} 
\begin{tabular}{ c c c c c c}
\hline\hline
& &MOND & CDM1& CDM2 & Newton  \\
\hline

Stars& $r_\beta =$ &$ 0.01$    & $0.01$& $0.01$  &    $0.01$    \\
  & $\beta_0 =$ &$ -0.5$    & $-0.5$& $-0.6$  &    $-0.5$    \\
  & $\beta_1 = $&$1$        & $1.2$& $1.1$  &    $1.$    \\

\hline
PN    & $r_\beta =$ &$ 2$    & $2$ & $2$  &    $2$    \\
  & $\beta_0 =$ &$ -1$    & $0.25$& $-0.15$  &    $-1$    \\
  & $\beta_1 = $&$2$    & $0.75$&  $ 1.25$   &    $1.8$    \\
\hline
Sat    & $r_\beta =$ &$ 20$    & $200$&  $200$  &    -    \\
  & $\beta_0 =$ &$ -1$    & $0.98$ &  $-0.5$  &    -    \\
  & $\beta_1 = $&$1.5$    & $-0.22$ &  $0.5$  &    -    \\

\hline
\end{tabular}
$\beta(r)=\beta_0+\beta_1 r/(r+r_\beta)$;    $r_\beta$ in kpc \\
\end{table} 

\vspace{-0.3 cm}
\subsection{MOND}
All scales are easy to fit with MOND.
\begin{itemize}
 \item Stars:
under $1$ kpc, the gravitational potential is purely Newtonian, so MOND and Newton models 
are identical. The anisotropy needs to be a bit radial ($\beta\sim 0.2-0.4$), but always far
from the allowed limits.
 \item PN:
until $5$ kpc the potential is still Newtonian. A slight difference appears after $8$ kpc. When varying 
$\beta(r)$ from $-0.6$ (tangential) to $0.8$ (radial), the MOND model is in good agreement with the observations. 
\item Satellites:
$\sigma_{los}^{sat}$ is also well-fitted by MOND on this scale, using $\beta=-0.2$ at $r=20$ kpc 
to $\beta=0.2$ at $r=200$ kpc.
\end{itemize}
 
\begin{figure}[ht!]
\center{
\includegraphics[width=8cm,angle=0]{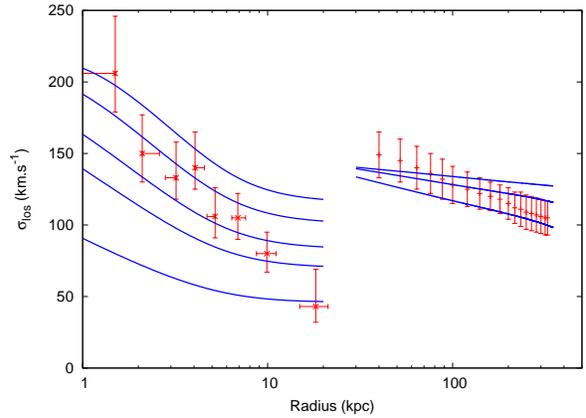}
\caption{ Fit of the velocity dispersion data around NGC 3379
with MOND, and various values of $\beta$. 
The data points are the same as in Fig. \ref{bestfit}.
 The 4 curves on a small-scale are $\beta$=-1, 0., 0.5, 0.9; and $\beta$=-1, 0.,0.5 on a large-scale.
}
\label{betafit}
}
\end{figure}

Figure \ref{betafit} shows how the variations in $\beta$ can help to fit many 
versions of the velocity curves. As in all other models, there is still much latitude to the fit. Large spatial variations of $\beta$ are possible if the ellipticals are the results of mergers (Dekel et al., 2005)

\vspace{-0.2 cm}
\section{Discussion}

In contrast to the conclusion of KP07, MOND does not predict constant
velocity dispersion with radius in the neighborhood of elliptical galaxies.
There is a wide latitude for varying the possible 
anisotropy parameter according to the  scale and the tracer considered. 
Our best fit starts from  a tangential or isotropic configuration near the center and
evolves progressively to a radial one for each of the three
scales considered, which appears quite realistic.

While the Newtonian model without dark matter has problems in the outer parts,
the CDM1 model encounters severe difficulties in the inner parts.
 The CDM2 model, with a reduced dark matter relative to the visible mass,
can also fit the data quite well.
Compared to the MOND model,  it requires a larger radial anisotropy on each scale,
and its $\beta$ profile  is unusual for the satellite tracer with a negative slope. For  $\beta$= constant,
the $\sigma_{los}^{sat}$ would be too steep compared to the observations. And by introducing anisotropy increasing with radius,  the $\sigma_{los}^{sat}$ slope is even more increased.

In the MOND regime, the external field effect (EFE, e.g. Wu et al 07) would also change
the predictions. This will modulate
the actual force on the particle tracers, and on the velocity.
Since there was still latitude for fitting the observations with MOND,
we feel that it is still possible to consider it
with other anisotropy parameters. However, we note that
the external field effect is not even necessary, for reproducing
the observed velocity dispersion slope.

%

\vspace{-0.2 cm}
{}


\begin{thebibliography}{}
\bibitem{} Angus G.W., Shan H.Y., Zhao H.S., Famaey B.: 2007a, ApJ, 654, L13 
\bibitem{} Angus, G.W., Famaey, B., Tiret, O., Combes, F., Zhao, H.S.: 2007b MNRAS, in press, arXiv/0709.1966   
\bibitem{} Bosma A.: 1981, AJ 86, 1825
\bibitem{} de Blok, W.J.G: 2005 ApJ  634, 227
\bibitem{} Dekel, A., Stoehr, F., Mamon, G.A. et al. : 2005 Nature, 437, 707
\bibitem{} Douglas, N.G., Napolitano, N.R., Romanowsky, A.J., et al.: 2007, astro-ph/0703047
\bibitem{} Famaey B., Bruneton J.P., Zhao H.S.: 2007 MNRAS 377, L79
\bibitem{} Gentile, G., Salucci, P., Klein, U. et al.: 2004 MNRAS, 351, 903
\bibitem{} Klypin A., Prada F.: 2007, astro-ph/0706.3554 (KP07)
\bibitem{} Milgrom M.: 1983, ApJ 270, 365
\bibitem{} Navarro, J.F.,  Steinmetz M.: 2000 ApJ 528, 607
\bibitem{} Romanowsky, A.J., Douglas, N.G., Arnaboldi, M. et al.:  2003, Science, 301, 1696
\bibitem{} Rubin, V.C., Thonnard, N., Ford, W.K., Jr.: 1980, ApJ 238, 471
\bibitem{} Sanders R.H., McGaugh S.S.: 2002, ARAA 40, 263
\bibitem{} Sanders R.H. : 2003 MNRAS, 342, 901  
\bibitem{} Shapiro K.L, Cappellari M., de Zeeuw T. et al.: 2006 MNRAS 370, 559
\bibitem{} Verheijen M.A, Sancisi R.: 2001, A\&A 370, 765
\bibitem{} Wu X., Zhao H.S., Famaey B. et al.: 2007 ApJ 665L, 101 
\end{thebibliography}
\end{document}